# Backward stimulated radiation from filaments in Nitrogen gas and air pumped by circularly polarized 800 nm femtosecond laser pulses

**Sergey Mitryukovskiy, Yi Liu\*, Pengji Ding, Aurélien Houard,
and André Mysyrowicz[+]**

*Laboratoire d'Optique Appliquée, ENSTA ParisTech/CNRS/Ecole Polytechnique, 828, Boulevard des Maréchaux, Palaiseau, F-91762, France*

\* *yi.liu@ensta-paristech.fr;* [+]*andre.mysyrowicz@ensta-paristech.fr*

Abstract: We report on strong backward stimulated emission at 337 nm in Nitrogen gas pumped by circularly polarized femtosecond laser pulses at 800 nm. A distinct dependence of the backward UV spectrum on pump laser polarization and intensity is observed, pointing to the occurrence of backward amplified spontaneous emission inside filaments. We attribute the population inversion to inelastic collision between the free electrons produced by the pump laser and neutral $N_2$ molecules. The addition of Oxygen molecules is detrimental for the gain, reducing it to near threshold at atmospheric concentration.

## 1. Introduction

Ambient air pumped by intense ultrashort laser pulses can give birth to stimulated radiation in both the forward and backward directions [1-12]. The backward stimulated radiation (propagating in the direction opposite to the pump laser) is of particular interest for remote sensing, and has therefore attracted considerable attention [1-6]. The advantage of the backward stimulated radiation for remote sensing lies in the fact that it can bring information about pollutants towards the ground observer with a well defined directionality. This has to be compared to the omni-directionality of the fluorescence or scattered optical signal from the same pollutants. In addition, if the information is carried to the ground observer via coherent Stokes- (SRS) or anti-stokes Raman scattering (CARS), the signal is proportional to $N^2$, as compared to the $N$ dependence of the fluorescence or spontaneous scattering process. Here $N$ is the number of pollutant molecules.

Up to now, two different schemes for backward stimulated radiation in air have been demonstrated experimentally, based on population inversion either of O atoms or $N_2$ molecules. In the first scheme, inversion in the population of Oxygen atoms was obtained with an intense UV pulses at 266 nm. The UV pulse served two functions, first to photo-dissociate Oxygen molecules, then to pump the Oxygen atoms to the $3p^3P$ state via two-photon excitation [1]. Strong stimulated emission at 845 nm, corresponding to the $3p^3P \rightarrow 3s^3S$ transition, was observed in both the backward and forward direction. A serious limitation of this scheme is the significant absorption of 226 nm light by atmosphere, preventing its use for remote pollutant sensing.

The second scheme deals with population inversion in neutral $N_2$ molecules. About 10 years ago, Q. Luo and coworkers announced the existence of a backward amplified spontaneous emission (ASE) from femtosecond laser filaments in air with *linearly* polarized 800 nm laser pulses. The evidence was based on the observation of a weak exponential increase of the backward UV luminescence at 357 nm upon filament length [5]. The UV emission corresponded to a transition between the $C^3\Pi_u$ ($v = 0$) and the $B^3\Pi_g$ ($v = 1$) triplet states of the N$_2$ molecules, where $v$ is the vibrational quantum number. Similar results have been confirmed recently by S. Owada and coworkers [6]. In 2012 D. Kartashov *et al.* reported a strong backward ASE lasing from N$_2$ molecules. A high power mid-infrared (3.9 μm or 1.06 μm) femtosecond pulses was used to induce a filament plasma column in a high pressure mixture of Nitrogen and Argon gas. Microjoule amounts of backward ASE from the plasma column were observed with a well-defined spatial profile [2]. In this experiment, the population inversion mechanism was the traditional Bennet scheme, where collisions transfer the excitation energy of Argon atoms to molecular Nitrogen. Unfortunately, this scheme cannot be employed for remote sensing application in atmospheric air, because of its requirement of high pressure Argon gas ($p > 3$ bar).

In this paper, we demonstrate that an intense backward stimulated radiation from filaments in Nitrogen gas can be achieved with *circularly* polarized femtosecond laser pulses at 800 nm. Filaments offer a favorable geometry for ASE because of the high aspect ratio of the plasma strings at the origin of the emission. Existence of the stimulated radiation is confirmed by the distinct dependence of the backward UV spectrum on the incident laser polarization and intensity. We attribute the population inversion between the triplet $C^3\Pi_u$ and $B^3\Pi_g$ states to inelastic collision between the electrons liberated by the pump laser and surrounding neutral ground state N$_2$ molecules. The dependence of the lasing effect on the incident laser pulse polarization is explained by the fact that a circularly polarized laser produces more energetic electrons than a linearly polarized one. We find that the presence of Oxygen molecules results in a significant quenching of the lasing action. With our present experimental parameters, significant gain occurs up to an Oxygen concentration of ~ 12%.

## 2. Experimental setup

In our experiments, femtosecond laser pulses with duration of 50 fs were focused by a convex lens of 1000 mm in a gas chamber filled with 1 bar of pure Nitrogen gas or a mixture of Nitrogen and Oxygen. A broadband dielectric beam splitter was used to steer the incident pulses into the gas chamber, while transmitting the backward UV emission from the gas plasma to the detector. A quarter-wave plate was inserted before the incident windows of the gas chamber to change the laser polarization between linear and circular. The backward emission was focused by a $f$ = 100 mm fused silica lens to the slit of a monochromator (Jobin-Yvon H-20 UV, grating: 1200 g/mm) combined with a photomultiplier tube (PMT). Typically, an average over 500 laser shots was performed for each individual measurement. The transverse fluorescence from the plasma channel was also measured. In this case a vertical slit of 1 mm was installed close to the center of the filament in order to limit the investigated plasma volume. The fluorescence transmitted through the slit was first collimated by a $f$ = 2.5 mm fused silica lens and then focused by another $f$ = 100 mm fused silica lens to the incident slit of the same monochromator and PMT (see Fig. 1).

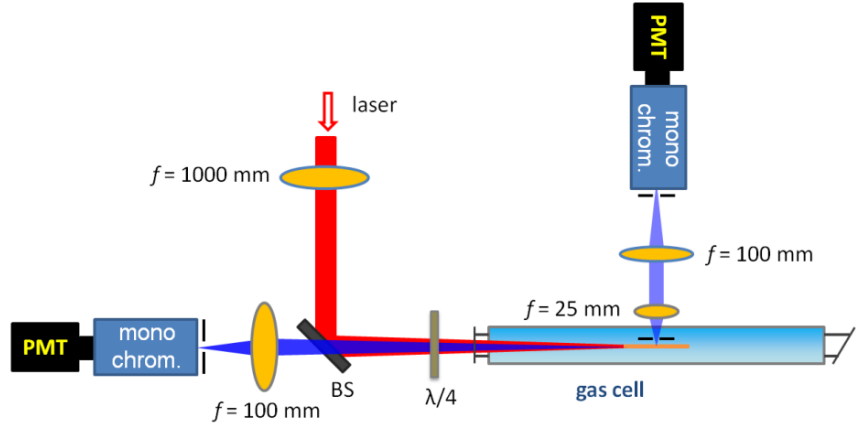

Fig. 1. Schematic experimental setup. The incident femtosecond pulse is focused by an $f$ = 1000 mm lens. The dichromatic beam splitter (BS) steers the 800 nm pulses inside the gas cell filled with Nitrogen gas or its mixture with air. The backward radiation from the plasma is collected by a $f$ = 100 mm fused silica lens into the slit of a monochromator. The signal is measured with a photomultiplier tube (PMT).

**3. Experimental results**

3.1. Spontaneous UV emission from filaments in Nitrogen gas

We first examine the spontaneous UV fluorescence emitted by filaments in a pure Nitrogen gas. The fluorescence spectra recorded in the transverse detection geometry are shown in Fig. 2 (a) for linearly and circularly polarized pump laser pulses. The emission peaks centered at 315, 337, 357, 380, 405 nm have been previously identified as due to transitions between various vibronic levels of the triplet $C^3\Pi_u$ and $B^3\Pi_g$ states of the neutral $N_2$ molecule, *i.e.* the second positive band of the $N_2$ molecules [2]. For all these five spectral lines, the signals are ~ 2 times stronger with circularly polarized laser than with linear laser polarization. A possible explanation will be discussed below.

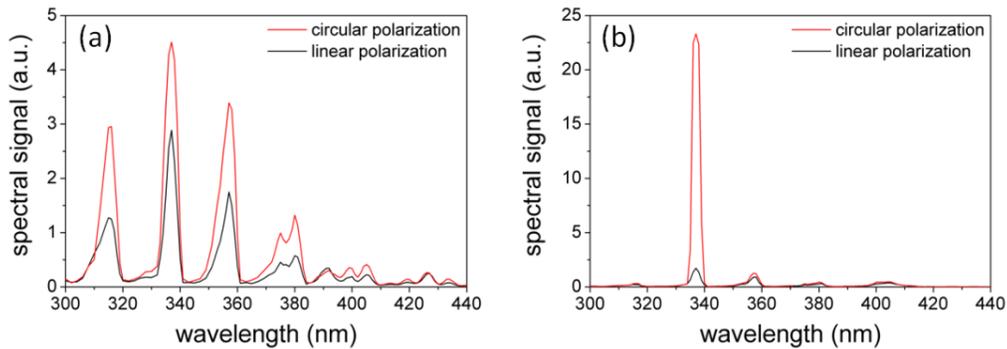

Fig. 2. Spectra of (a) transverse fluorescence and (b) backward UV emission for circular and linear laser polarization. The incident laser pulse energy is 9.3 mJ.

Before discussing the backward emission at 337 nm with circularly polarized laser, we first examine in more details the results obtained with linear laser polarization in the backward direction, a geometry where ASE was reported previously [5, 6]. In Fig. 3 (a), we present the measured backward 337 nm signal as a function of the incident laser pulse energy up to 9.3 mJ. We also measured the length and the width of the plasma channel (defined at $1/e^2$ level) as a function of laser pulse energy (see Fig. 3 (b)). With an increase of the laser energy, the filament length increases linearly, in agreement with a recent observation [6].

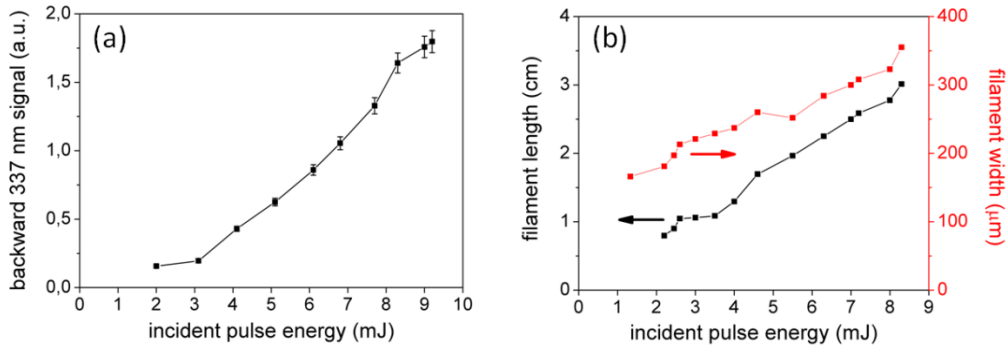

Fig. 3. (a) Backward 337 nm signal as a function of incident laser pulse energy. (b) Measured length and width of plasma as a function of pump laser energy. In both cases the pump laser is linearly polarized.

In Fig. 4, we plot the 337 nm signal as a function of the filament length, as it was previously done for identification of ASE [5, 6]. A weak nonlinear dependence is observed, which is tentatively fitted with an exponential law. We note that a more satisfying procedure consists in plotting the signal as a function of plasma volume, in view of the fact that the filament size increases slightly with pump laser energy (see Fig. 3 (b)). A plot of the backward signal as a function of plasma volume is best fitted by a linear dependence, as shown in Fig. 4. Therefore, we come to the conclusion that stimulated radiation at 337 nm is not achieved with linearly polarized femtosecond pulses in our experiments. We also performed measurements at 357 nm and observed a similar dependence.

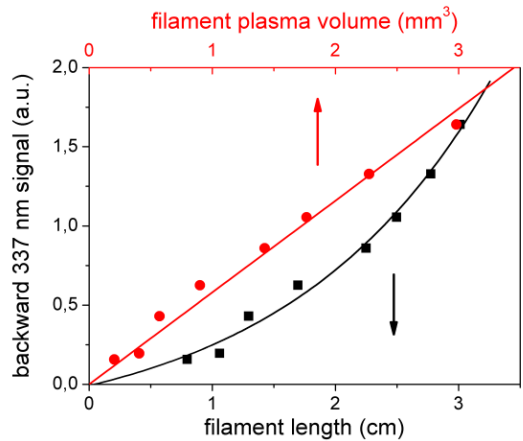

Fig. 4. Backward 337 nm signal as a function of plasma length (black squares, lower scale) and plasma volume (red circles, upper scale). The curves are best fitted with an exponential or linear law respectively.

3.2. Stimulated UV emission from filaments in Nitrogen gas

We now concentrate on the backward emission at 337 nm obtained with circularly polarized femtosecond laser pulses. It corresponds to the (0-0) vibronic transition of the second positive band system of $N_2$ molecule. In Fig. 2 (b), the spectra of the backward UV emission are shown for circular and linear polarization of the laser. The emission intensity at 337 nm is now about 10 times larger with circular pump laser polarization than with linear polarization.

Others lines at 315, 357, 380, and 405 nm increase by a factor of ~1.5, an increase similar to that of the spontaneous fluorescence presented in Fig. 2 (a). The remarkable behavior of the 337 nm signal suggests that backward stimulated emission is initiated for this particular line with circularly polarized laser pulses.

In Fig. 5, the backward emission intensity at 337 nm is plotted as a function of incident laser energy. The 337 nm signal displays a superlinear dependence on incident laser energy.

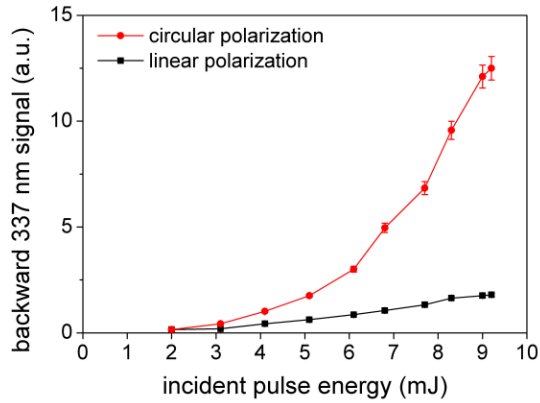

Fig. 5. Measured backward signal at 337 nm as a function of incident laser pulse energy, for both circular and linear laser polarization.

To get further insight into the distinct pump polarization dependence presented in Fig. 5, we measured the backward radiation intensity at 337 nm as a function of the incident laser ellipticity. In Fig. 6, the measured signals are presented as a function of the rotation angle of the quarter-wave plate for different incident laser energies. The angles $\varphi = 90° \times m$ correspond to linearly polarized laser, with $m$ = 0, 1, 2, 3. The angles $\varphi = 45° + 90° \times m$ correspond to circularly polarized laser. For a low pulse energy of 300 µJ, linearly polarized pulses generates a UV radiation with an intensity twice that obtained with circular polarization (Fig. 6 (a)). For an increased incident energy of 660 µJ, an octagon-shaped dependence is observed (Fig. 6 (b)), indicating the onset of a new mechanism for the emission at 337 nm. In the case of $E_{in}$ = 9.3 mJ, the signal obtained with circular laser polarization totally dominates and decreases rapidly when the laser polarization deviates slightly from circular. This critical dependence of the emission signal at 337 nm with laser polarization reinforces the hypothesis that backward stimulated emission at 337 nm occurs inside the filament plasma.

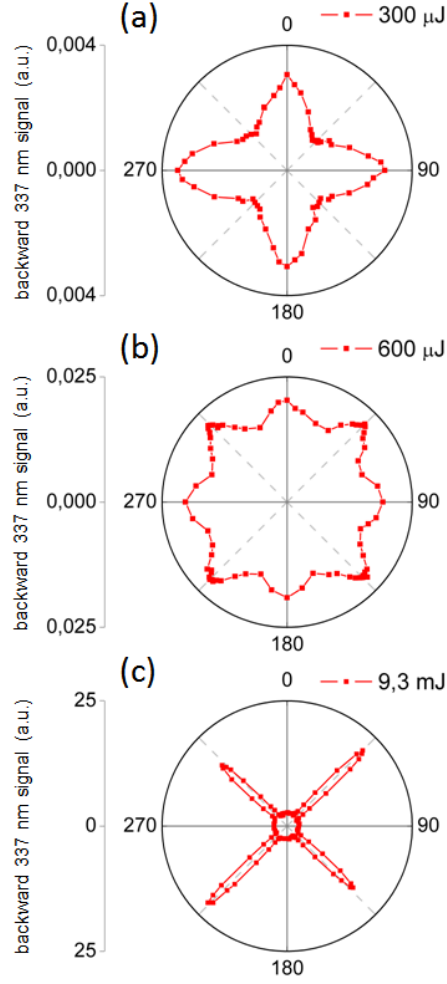

Fig. 6. Backward emission signal at 337 nm as a function of the rotation angle of the quarter-wave plate. The incident laser energy for (a), (b), (c) are 0.3 mJ, 0.6 mJ, and 9.3 mJ, respectively.    Angle 0º corresponds to linearly polarized light.

## 4. Discussion of results

We now discuss possible mechanisms for population inversion between the $C^3\Pi_u$ and $B^3\Pi_g$ states of the neutral $N_2$ molecule, at the origin of the backward stimulated radiation. First, it is worth reminding that a direct population transition between the ground singlet state $X^1\Sigma_g^+$ and the excited triplet $C^3\Pi_u^+$ state is forbidden in the electric dipole approximation. One widely discussed mechanism to populate $C^3\Pi_u^+$ state inside filaments in air is the following reaction [13]:

$$N_2^+ + N_2 = N_4^+,$$
$$N_4^+ + e = N_2^+(C^3\Pi_u^+) + N_2(X^1\Sigma_g^+).$$

With this mechanism, one expects that a linearly polarized laser produces a stronger fluorescence signal for two reasons. First, it is known that the ionization rate of atoms and molecules is higher for linear polarization both in the multiple photon ionization and tunneling ionization regime [14]. As a result, the densities of electrons and positive ions $N_2^+$ are higher in the case of linear laser polarization and should result in a higher density of $N_2^+(C^3\Pi_u^+)$. Second, the nonlinear refractive index of ambient air $n_2$ is higher for linear polarized laser [14]. This results in slightly higher laser intensity inside filaments, which should again lead to higher densities of electrons and ions. We therefore assume that this collision-assisted recombination process of the electron on the parent ion is responsible for the fluorescence of filaments at relatively low laser energy, such as that of $E_{in}$ = 300 μJ in Fig. 6 (a), but not to the stimulated emission observed at higher pump powers.

Another mechanism for the transition from the ground state to the $C^3\Pi_u^+$ state is the electron-molecule inelastic collision:

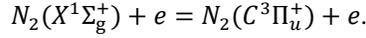

$$N_2(X^1\Sigma_g^+) + e = N_2(C^3\Pi_u^+) + e.$$

This is actually the main reaction responsible for population inversion in a traditional $N_2$ laser, where the electrons are accelerated to obtain sufficient energy by the discharge electric field [15]. The cross section of the above reaction is sensitive to the kinetic energy of the incident electron. It is nearly zero for electron energy below the threshold kinetic energy $E_{th}$ ~ 10 eV, exhibits a resonant peak around 14.1 eV, and then decreases progressively for higher energy electrons [16]. For electrons born in the intense laser field inside filaments, the distribution of kinetic energies depends strongly upon pump laser polarization. With linear polarization, free electrons are accelerated back and forth along the laser polarization direction at each optical cycle, so that at the end of the pulse most of them have a low kinetic energy, with a weak tail distribution extending to a few eV. For circularly polarized light, where electrons are always accelerated away from the ion core, the final electron distribution is nearly mono-kinetic with a peak at twice the ponderomotive energy. Calculations performed for a filament show that the peak of this distribution is between 5 and 7 eV assuming a laser intensity of $5\times10^{13}$ W/cm² and should scale up linearly with laser intensity [17]. The widely quoted clamped intensity value in air filaments of $5\times10^{13}$ W/cm² is therefore insufficient by a factor 2 to be an effective excitation mechanism. On the other hand, it is now well established that intensity clamping is not a rigorous concept. Intensity spikes exceeding the clamped value by order of magnitude have been measured in filaments [18].

In our current experiment, we estimated the laser intensity inside the filament by inserting an Al foil around the middle of the filament in the case of $E_{in}$ = 9.3 mJ, in ambient air. After several laser shots, a micrometric hole was drilled by the filament itself [19]. After 300 laser shots the transverse size of the hole became stable and its diameter (d) was measured to be 170 μm. At the same time, the transmitted laser energy $E_t$ was measured to be about 6.6 mJ. The average laser intensity can be estimated as:

$$I = \frac{E_t}{\pi\left(\frac{d}{2}\right)^2 \tau_p} = 1.45\times 10^{14} \text{ W/cm}^2,$$

where $\tau_p$ is the initial laser pulse duration of 50 fs. This indicates that the electron energy required for impact excitation of neutral $N_2$ molecules is well reached in filaments. More detailed measurements of the intensity inside filaments will be published elsewhere [19]. For linearly polarized laser field, this mechanism is turned off because of the low electron kinetic energy, even with a laser intensity above the clamped value. As a conclusion, we believe that the inversion of population inside the triplet manifold of $N_2$ is due to impact excitation by electron collisions. The threshold character for the inelastic collision cross section with respect to electron energy is reflected in the sharp dependence of the lasing emission with ellipticity [16].

## 5. Backward emission from filaments in ambient air

For applications of this backward stimulated radiation source for remote sensing, operation in ambient air is required. We therefore performed measurements in atmospheric air. We first examined the backward UV emission from filaments formed in ambient air driven by both circularly and linearly polarized laser pulses. Emission spectra similar to those of Fig. 2 (b) were observed, except that the signal at 337 nm increased only by a factor 2 for circular laser polarization. In Fig. 7, the measured backward 337 nm signal is shown as a function of the rotation angle of the quarter-wave plate.

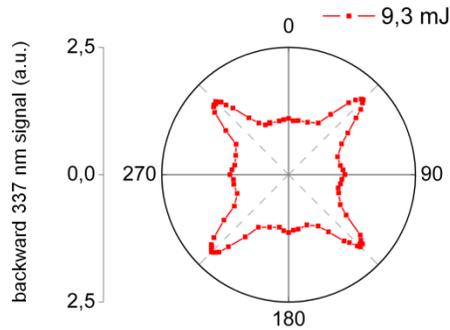

Fig. 7. Dependence of the backward 337 nm signal obtained from filaments in air as a function of the rotation angle of the quarter-wave plate. The incident laser energy is 9.3 mJ and the focal length is $f = 1000$ mm.

To further assess the influence of Oxygen gas for the lasing action, we measured the backward 337 nm signal in different mixtures of pure Nitrogen and Oxygen. We present in Fig. 8 the measured 337 nm signal for both circularly and linearly polarized laser as a function of the percentage of Oxygen. A slow decrease of the signal is observed upon increasing Oxygen concentration up to 10 %. Beyond 10 % Oxygen, the signal shows a rapid decrease, indicating the termination of significant lasing action. For Oxygen concentration more than 13%, the signal obtained with circular laser polarization becomes ~ 2 times larger than that of linear laser polarization, similar to the transverse fluorescence presented in Fig. 2 (a).

The detrimental influence of Oxygen molecule for the conventional discharge-pumped Nitrogen laser is well documented [15]. The underlying physical mechanism is attributed to the collision reaction

$$N_2(C^3\Pi_u^+) + O_2 = N_2(X^1\Sigma_g^+) + O + O,$$

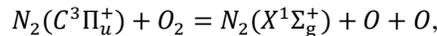

which efficiently reduces the population density in the upper state of the lasing emission. To obtain a significant backward stimulated emission in atmospheric air, a higher population inversion density (or a higher gain) is required to overcome the quenching effect of the Oxygen molecules. A possible approach to achieve this is to use a pump laser at longer wavelengths, because the kinetic energy of the produced electrons increases like $I\lambda^2$, where $I$ is the laser intensity and $\lambda$ its wavelength.

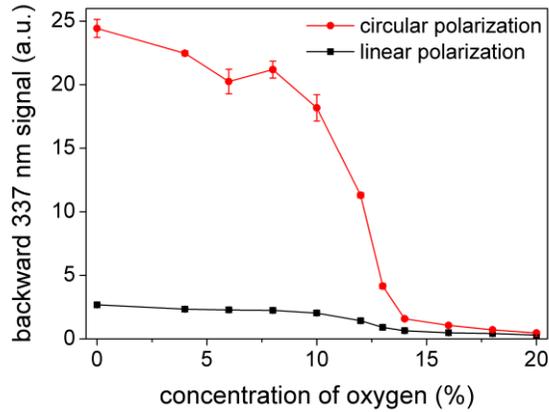

Fig. 8. Measured backward 337 nm signal from filaments in air as a function of Oxygen gas concentration, for both circular and linear polarized laser pulses. The incident pulse energy is 9.3 mJ.

## 6. Conclusion

In summary, we have shown that a strong stimulated radiation at 337 nm can be achieved in the backward direction from filament plasma in a Nitrogen gas pumped by a circularly polarized laser pulse at 800 nm. This stimulated radiation shows distinct dependence on laser pulse energy, compared to that obtained with linear polarized laser pulses. In a mixture of Nitrogen gas and Oxygen, the presence of Oxygen molecules suppresses the lasing action to a large extend. As to the mechanism responsible for population inversion, we attribute it to inelastic collisions between electrons liberated by the pump laser and neutral Nitrogen molecules, a process which is more efficient with circularly polarized laser pulses. We believe that this simple scheme for backward stimulated emission from Nitrogen gas pumped by the widely available 800 nm femtosecond laser pulse is a significant step towards applications for remote sensing.

## 7. Acknowledgments

We gratefully acknowledge useful discussions with P. Corkum of Ottawa University.